\documentstyle[aps,prb,epsf,epsfig]{revtex}
\begin{document}

\twocolumn[\hsize\textwidth\columnwidth\hsize\csname@twocolumnfalse\endcsname

\title{Nature of 45$^{\circ }$ vortex lattice reorientation in tetragonal
superconductors }
\author{A. Knigavko$^{a}$, V.G. Kogan$^{b}$, B. Rosenstein,$^{a,c}$ and T.-J. Yang$%
^{a}$}
\address{$^{a}$Department of Electrophysics, National Chiao Tung
University, Hsinchu, Taiwan 30050, Republic of China\\
$^{b}$Ames Laboratory and Physics Department, Iowa State University, Ames,
Iowa 50011, USA\\
$^{c}$National Center for Theoretical Sciences, National Chiao Tung
University, Hsinchu, Taiwan 30050, Republic of China}
\maketitle

\begin{abstract}
The transformation of the vortex lattice in a tetragonal superconductor
which consists of its 45$^{\circ }$ reorientation relative to the crys tal
axes is studied using the nonlocal London model. It is shown that the
reorientaton occurs as two successive second order (continuous) phase
transitions. The transition magnetic fields are calculated for a range of
parameters relevant for borocarbide superconductors in which the
reorientation has been observed.
\end{abstract}
\bigskip

]

Properties of the vortex matter have recently attracted great attention due
to diversity of phases and novel phenomena associated with them. One of the
main research goals is determination of the phase diagram. In high
temperature superconductors the vortex matter phases include the vortex
liquid and various vortex solids which exist due to the competition of
intervortex interactions with fluctuations both thermal and those due to the
quenched disorder. \cite{Blatter} On the other hand, in borocarbide
superconductors a rich variety of quite perfect vortex crystals has been
observed. The experimental information comes from such different
measurements as neutron diffraction, decoration, and scanning tunneling
microscopy.\cite{dec,neut,stm} For these near isotropic materials the
entropy contribution to the free energy is small and phase transitions in
the vortex lattice are governed by competition between intervortex
interactions of different symmetry.

The borocarbides are materials of the tetragonal symmetry. Interactions of
this symmetry should exist for any physical subsystem of the crystal. In
particular, in the mixed state with the field along the fourfold tetragonal
axis they would favor a square vortex lattice. However, the standard
magnetic repulsion of vortices is isotropic in this case. The isotropic
interaction becomes dominant when the intervortex distance is large enough
and a sparse lattice is close to hexagonal, the most closely packed two
dimensional lattice. One, therefore, expects that the interplay of the
interactions of different symmetries may result in structural
transformations of the vortex lattices, observed in borocarbides.

For the applied magnetic field along the fourfold symmetry axis, these
transformations are as follows. With decreasing magnetic field, the lattice
undergoes a second order phase transition, at which the square structure
loses stability and becomes a rhombic (distorted hexagonal) vortex lattice. 
\cite{neut,stm} As the field further decreases, the rhombic lattice changes
the orientation relative to underlying crystal by 45$^{\circ }$, which has
been classified as the first order transition.\cite{McPaul,dec,boro} For the
field along the twofold axis, the 90$^{\circ }$ reorientation has been
reported.\cite{PT-perp-field}

In this paper we study in detail the 45$^{\circ }$ reorientation and clarify
its nature. We show that this reorientation proceeds as two successive
second order (continuous) transitions and not as an abrupt first order
transition, the scenario assumed before. Instead of considering a limited
class of rhombic lattices,\cite{boro} we study the general class of
arbitrary lattices. We find that in the field region between the two second
order phase transitions, the lattice with the lowest possible symmetry is
realized (with the inversion being the only symmetry element). This
intermediate region is quite narrow and the structural evolution in this
field domain might be difficult to discern experimentally. However, the
thermodynamic characteristics of the superconductor are different for the
two scenarios, and this can be tested. In particular, no latent heat is
expected during the lattice reorientation. We also predict a peak in the
critical current in the transition region if the pinning is of a weak
collective type. Below we describe the London model with nonlocal
corrections relevant for the mixed state of borocarbides. Then, the
numerical procedure is outlined and the results are presented

A fruitful approach to the problem of the vortex lattice phases is the
extended London model. We start here with London equations corrected for
nonlocality:\cite{Kogan1,boro} 
\begin{eqnarray}
\frac{4\pi }{c}j_{i}({\bf k}) &=&-\frac{1}{\lambda ^{2}}q_{ij}({\bf k})a_{j}(%
{\bf k)}  \nonumber \\
&=&-\frac{1}{\lambda ^{2}}\left( m_{ij}^{-1}-\lambda
^{2}n_{ijlm}k_{l}k_{m}\right) a_{j}({\bf k)}.  \label{Extented-London}
\end{eqnarray}
Here, $a_{j}=A_{j}+(\Phi _{0}/2\pi )\partial _{j}\theta ,$ $A_{j}$ is the
vector potential, $\theta $\ is the order parameter phase, and $\Phi _{0}$\
is the flux quantum. The nonlocal response kernel $q_{ij}({\bf k})$\ is
expanded up to the second order terms in the wave vector ${\bf k}$. The
tensor $n_{ijlm}\propto \langle v_{i}v_{j}v_{l}v_{m}\rangle \gamma (T,\ell )$
where ${\bf v}$ is the Fermi velocity and the function $\gamma $ decreases
somewhat with temperature and drops fast for short mean-free paths $\ell $. 
\cite{Kogan2} It is difficult to accurately estimate the components of ${%
\hat{n}}$ because of uncertainties in determination of Fermi velocities and,
in particular, of the mean-free path. At low temperatures, ${\hat{n}}\sim
\gamma /\kappa ^{2}$, where $\kappa $ is the Ginzburg--Landau parameter.
Since good crystals of borocarbides are clean materials with $\kappa =10\div
15$, one expects the components of ${\hat{n}}$ to be of the order $10^{-2}$.
Note also that for the problem of vortex lattices in fields well under the
upper critical field, the correction $\lambda ^{2}{\hat{n}}k^{2}\sim \xi
_{0}^{2}k^{2}\ll 1$ ($\xi _{0}$ is the zero-$T$ coherence length).
Therefore, for strong type-II superconductors, the corrections to the
standard London equations and the truncation in the expansion (\ref
{Extented-London}) are well justified.

For the tetragonal symmetry, the tensor  ${\hat{n}}$ in the crystal frame
has four independent components $n_{xxxx}$, $n_{xxyy}$, $n_{zzzz}$, and $%
n_{xxzz}$ The inverse mass tensor has two different components $%
m_{xx}^{-1}=m_{yy}^{-1}$ and $m_{zz}^{-1}$. The London free energy
functional corresponding to Eq. (\ref{Extented-London}) reads 
\begin{equation}
F=\frac{1}{8\pi }\int \frac{{\rm d}{\bf k}}{4\pi ^{2}}\left( |{\bf h|}%
^{2}-\lambda ^{2}\epsilon _{ijk}\epsilon
_{lmn}k_{j}k_{m}q_{lk}^{-1}h_{n}h_{i} \right)  \label{Energy-general}
\end{equation}
where ${\bf h}({\bf k})$\ is the magnetic field and $\epsilon _{ijk}$ is the
unit antisymmetric tensor. The nonlocal corrections preserve linearity of
the London equations and do not change the standard London result that the
interaction of two vortices is proportional to the field of one of them at
the location of the other.\cite{Kogan2} As usual, the free energy density of
a vortex lattice is given by $F=\left( B^{2}/8\pi \Phi _{0}\right) \sum_{%
{\bf g}}h_{z}({\bf g})$ where $B$ is the magnetic induction, ${\bf g}$\ is a
vector of the reciprocal lattice and $h_{z}$ is the component of the single
vortex field along the vortex axes. We are interested in the field along the
fourfold symmetry axis $z$. Solving Eq. (\ref{Extented-London}) for a single
vortex one can bring the free energy density to the form\cite{boro} 
\begin{equation}
F=\sum_{{\bf g}}\frac{B^{2}/8\pi }{1+\lambda ^{2}g^{2}+\lambda
^{4}(n\,g^{4}+d\,g_{x}^{2}g_{y}^{2})}\,,  \label{Energy-VL}
\end{equation}
where $n=n_{xxyy}$ and $d=2(n_{xxxx}-3n_{xxyy}).$ The free energy $F(B,T)$\
is the thermodynamic potential, which is minimum in equilibrium of a
superconducting slab in a perpendicular applied field. The temperature
enters $F(B,T)$ via $T$ dependent parameters $\lambda (T),$\ $n(T)$\ and $%
d(T)$\ that can, in principle, be calculated using a microscopic model. Note
that besides the factor $B^{2}$, the induction enters via the area of the
primitive lattice cell. We determine the stable lattice by numerical
minimization of $F(B,T;{\bf g})$ with respect to the lattice structure
specified by a given set of ${\bf g}$'s.

\begin{figure}[htp]
\epsfig{figure=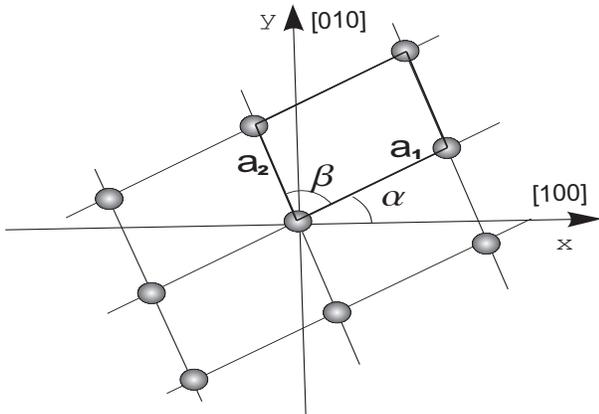,height=5.5cm,width=8cm}
\caption{
General vortex lattice and its orientation relative to the
crystal.
 }
\end{figure}

The vortex lattice is completely defined by the basis vectors ${\bf a}_{1}$\
and ${\bf a}_{2}$, i.e., by four parameters. Since a unit cell accommodates
one flux quantum, $a_{1}a_{2}\sin \beta =\Phi _{0}/B$, three parameters
suffice. Following Ref.\onlinecite {Saint-James} we choose $\alpha $, $\rho
\equiv (a_{2}/a_{1})\cos \beta $, and $\sigma \equiv (a_{2}/a_{1})\sin \beta 
$ as the needed three (see Fig. 1 for definitions of $\alpha $ and $\beta $%
). The parameters $\rho $ and $\sigma $ are convenient because one can
select a domain of their variation, each point of which corresponds to a
lattice with various equivalent choices of the basis vectors ${\bf a}_{1,2}$%
. Thus, the minimization of $F$ is done at fixed $B$, $n$, and $d$\ \ with
respect to \ $\rho ,\sigma $\ and $\alpha $ for $0<\alpha <\pi $, $0\leq
\rho \leq 0.5,$ and $\rho ^{2}+\sigma ^{2}\geq 1$.\cite{Saint-James} The
minima of $F$ are often located on the boundaries of this domain; we use the
``Amoeba'' numerical routine convenient in such circumstances.\cite{recipes}
The cutoff factor $\exp \left( -\xi ^{2}g^{2}\right) $ was introduced inside
the sum (\ref{Energy-VL}) to properly account for the failure of the London
model in the vortex core. Changing parameters $B$\ and $n,d$ we obtain the
phase diagram.

\begin{figure}[tph]
\epsfig{figure=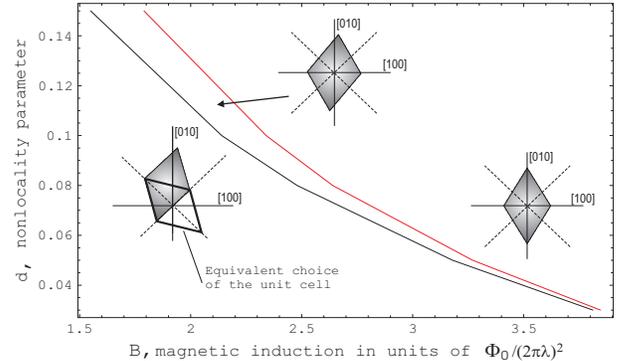,height=4.8cm,width=8cm}
\caption{
 Phase diagram of the vortex lattice in the region of the
reorientation for $n=0.015$. Nonlocal parameters $n$ and $d$ are defined in
the text. The magnetic induction $b$ is in units of $\Phi_0/(2\pi\lambda)^2$.
}
\end{figure}

The main finding of this work is that the reorientation of the lattice
proceeds as two steps. Figure 2 shows the transition lines on the $B,d$
plane for a fixed $n=0.015$. The equilibrium lattices both before and after
the reorientation have the rhombic symmetry $D_{2h}$.\cite{Note} Their
symmetry axes, which coincide with the diagonals of a rhombic unit cell
(with the appropriate choice of such a cell, see Fig. 2), are aligned with $%
[110]$\ and $[1\bar{1}0]$ at lower magnetic inductions, whereas the symmetry
axes are at $[100]$\ and $[0\bar{1}0]$\ for higher $B$'s. This result is in
accordance with data for $YNi_{2}B_{2}C$.\cite{McPaul} In a narrow region
between the two rhombic phases, a less symmetric lattice is stable. Here,
the unit cell is a general parallelogram. All in-plane symmetry elements
disappear except the inversion, and the symmetry group reduces to $C_{2h}.$\

One could describe the reorientation process as a gradual rotation of the
unit cell accompanied by a slight deformation. Figure 3 shows how the angles 
$\alpha $\ and $\beta $\ change when $B$\ increases in the vicinity of the
reorientation (for $d=0.05$ and $n=0.015$). The transitions at $B\approx
3.18\,\Phi_0/(2\pi\lambda)^2$ and $B\approx 3.27\,\Phi_0/(2\pi\lambda)^2$
are seen clearly. The angles $\alpha$, $\beta$, and the other lattice
parameters are continuous at the two transition fields. We conclude that
both phase transitions that occur during the reorientation are of the second
order. The sequence of symmetry changes with the field decreasing is $%
D_{2h}\rightarrow C_{2h}\rightarrow D_{2h}.$\ While at the first step of the
reorientation, the symmetry becomes lower, at the second step the symmetry
increases. Correspondingly, the ground state is double degenerate in both $%
D_{2h}$ phases; the degenerate vacua (two equilibrium structures of the same
energy) are related by rotations over $90^{\circ }$). The structure becomes
four times degenerate in the intermediate $C_{2h}$ phase (rotations over $%
\pm 45^{\circ }$ and $90^{\circ }$). Practically, this may lead to
apparently increased disorder in the $C_{2h}$ phase.

\begin{figure}[htp]
\epsfig{figure=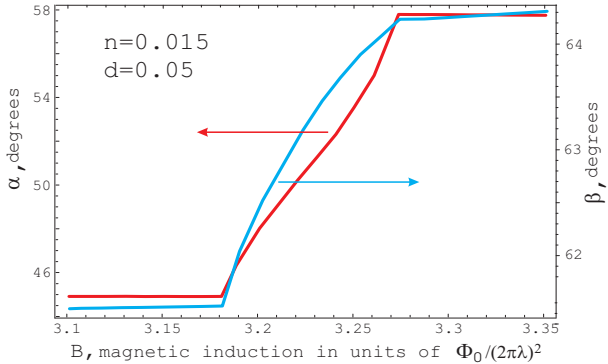,height=4.8cm,width=8cm}
\caption{
 Evolution of angles $\alpha $\ and $\beta $ (defined in Fig. 
1) with  field $b=\Phi_0/(2\pi\lambda)^2$ during reorientation.\\
 }
\end{figure}

It is worth noting that the relative energy differences between the
equilibrium $C_{2h}$ lattice and the rhombic ones is exceedingly small. As
an example, we provide this figure for $d=0.05$ and $n=0.015$: the relative
difference between energies of the rhombic lattice at the transition point
and the lattice in the middle of the field domain of $C_{2h}$ structure is
of the order $10^{-7}$. This is much smaller than $10^{-2}$ for the relative
energy differences usually cited for triangular and square lattices within
the standard London or Ginzburg--Landau models.

The location of the phase transition lines is sensitive to both $n$ (for the
isotropic correction in Eq. (\ref{Energy-VL})) and $d$ (of the four-fold
symmetric correction). Figure 4 shows the phase diagram of the vortex
lattice on $B,d$ and $B,n$ planes in the region of the reorientation
process. The region of stability of the monoclinic lattice is broader for
smaller $n$'s and larger $d$'s. Still, as is seen at Fig. 4, this field
region is narrow for values of $n$ and $d$ adopted in our simulations. For
example, for {\sl LuNi}$_{2}${\sl B}$_{2}${\sl C } with $\lambda \approx
710\,$\AA , the field unit $\Phi _{0}/(2\pi \lambda )^{2}$ is about $100\,$G.

The new scenario of the lattice reorientation in a tetragonal superconductor
which is found this paper has implications for thermodynamic characteristics
of the vortex lattice and its dynamic behavior. We have found, and this is
our main result, that the reorientation proceeds as two successive phase
transitions of the second order when the applied field or temperature vary.
Therefore, continuous variation of the entropy (i.e., no latent heat) and of
the reversible magnetization are expected during reorientation. In contrast,
the old scenario of the first order transition implied discontinuous jumps
of the above quantities.

As is seen at Fig. 4, for small values of $n$ and $d$, the domain of
monoclinic phase shrinks. Then, it would be difficult to distinguish
experimentally this situation from 1st order transition, because the entropy
would change fast with $B$ during the reorientation (for the $B$-sweep at a
fixed $T$). Still, one should not observe hysteresis, characteristic of the
1st order transitions. If the sequence of transitions we suggest here is
found, it would be of interest to suppress $n$ and $d$ by making the
mean-free path shorter and to see how the transition evolves (as has been
done with doping Lu-based borocarbide crystals with Co.\cite{Co,Bell})

\begin{figure}[tph]
\epsfig{figure=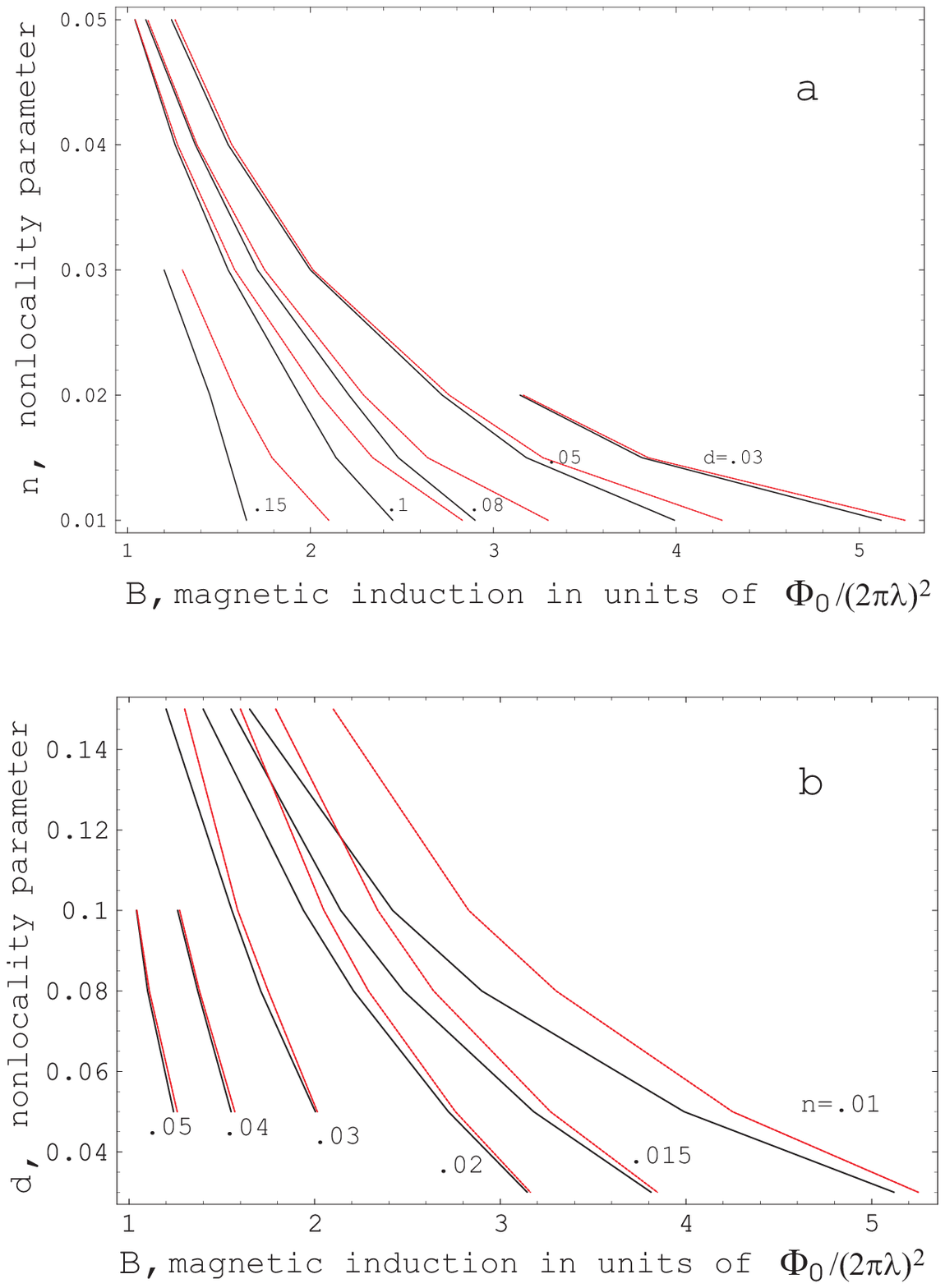,height=9.5cm,width=8cm}
\caption{
Phase diagram of the reorientation transformation (a) in the $n,b$
plane for a set of $d$'s indicated, and (b) in the $d,b$ plane for a set of $%
n$'s indicated.
 }
\end{figure}

Both the upper and lower phase transitions ($D_{2h}\leftrightarrow C_{2h}$)
of the reorientation process cause uniform spontaneous deformations of the
vortex lattice. As a result, a particular combination of the elastic lattice
moduli vanishes at the transitions. It has been recently shown that a change
of elastic properties of this type leads generally to peculiarities in the
critical current, provided a weak collective pinning operates in the
material.\cite{Rosen} Therefore, the reorientation of the vortex lattice in
borocarbides may lead to a peak in the critical current.

Finally, we would like to point to other possible applications of our
results. The London model we employed reflects properly the symmetry of the
system. It was originally derived for an anisotropic Fermi surface and
isotropic superconducting pairing.\cite{Kogan1} However, the d-wave type of
pairing also leads to a similar effective London model\cite{Affleck} (for
not very low temperature where the effects of the order parameter nodes
become essential). The reorientation of the vortex lattice has indeed been
found theoretically, and characterized as the first order transition.\cite
{Franz} It would be of interest to check whether or not our scenario of the
reorientation applies to this case as well.

This work is supported by NSC of Taiwan through the grants
\#89-2112-M-009-0016 and \#89-2112-M-009-039.


\begin{references}
\bibitem{Note}  To avoid misunderstanding we point out that in all cases we
speak about the symmetry of gauge invariant quantities.
\end{references}
\end{document}